\documentclass{article}
\usepackage{latexsym}

\def\p{\mbox{\boldmath $p$}}

\def\D{{\cal D}}  
\def\H{{\cal H}}  
\def\L{{\cal L}}  
\newtheorem{theorem}{Theorem}[section]
\newtheorem{lemma}[theorem]{Lemma}
\newtheorem{corollary}[theorem]{Corollary}
\newtheorem{proposition}[theorem]{Proposition}

\newenvironment{proof}{\noindent {\bf Proof\,\ }}{\hfill\mbox{\ $\Box$} \smallskip}

\begin{document}

\title{Notes on computing peaks in k-levels and 
parametric spanning trees}
\author
{Naoki Katoh\thanks{Department of Architecture, Kyoto University,
Kyoto, 606-8501, Japan. 
Email:naoki@archi.kyoto-u.ac.jp}
\and Takeshi Tokuyama\thanks
{Graduate School of Information Sciences, Tohoku University,
Sendai, 980-8579, Japan. Email:tokuyama@dais.is.tohoku.ac.jp}
}

\maketitle

\begin{abstract}

We give an algorithm to compute all the local peaks in 
the $k$-level of an arrangement of $n$ lines in 
$O(n \log n) + \tilde{O}((kn)^{2/3})$ time.
We can also find $\tau$ largest peaks in 
$O(n \log ^2 n) + \tilde{O}((\tau n)^{2/3})$ time.
Moreover, we consider the longest edge in a parametric
minimum spanning tree (in other words, a bottleneck edge 
for connectivity), and  
give an algorithm to compute the parameter 
value (within a given interval) 
maximizing/minimizing the length of the longest edge in MST.
The time complexity is $\tilde{O}( n^{8/7}k^{1/7} + n k^{1/3})$. 
\end{abstract}

\section{Introduction}

The $k$-level of an arrangement of lines is one of 
popular geometric objects in 
computational geometry~\cite{Edels,WEB}.
The $k$-level is the union of 
$k$-th lowest (closed) line-segments of the arrangement, 
and it can be considered as the trajectory 
of the $k$-th smallest element in a set of $n$ data 
each of which depends on a parameter $x$ linearly.
Thus, the $k$-level is a special case of the locus of 
the largest element of the minimum base of 
a parametric matroid~\cite{G79,G80,Eppstein95,KIT95,KTT99} with 
one parameter; 
in precise, the $k$-level is the locus of the 
maximum element in the minimum base of a parametric uniform
matroid of rank $k$.  
From a different aspect, 
the $k$-level is a dual concept of the $k$-set~\cite{Edels}, 
and  the complexity $g_k(n)$ of the $k$-level
of an arrangement of $n$ lines is asymptotically same 
as the number of different $k$-sets in a set of $n$ points
in a plane. 

Lov\'{a}sz~\cite{L71} first gave a nontrivial 
$O(n^{3/2})$ upper bound for $g_{\lfloor n/2 \rfloor}(n)$, 
and also 
introduced Straus's $\Omega( n \log n )$ lower bound.  
The current best upper and lower bounds for $g_k(n)$ are
$O( k^{1/3} n)$~\cite{Dey97} and 
$2^{\Omega(\sqrt{\log k})} n$ ~\cite{Toth00}, respectively.
The upper bound holds for 
any parametric matroid (with a linear parameter) of 
rank $k$ in $n$ elements~\cite{Eppstein95}.  
Moreover, Cole et al. gave 
an $O(g_k(n) \log ^2 n)$~\cite{CSY87} time algorithm to
compute the $k$-level; this time complexity has been 
improved due to improvement of 
algoirthms for the dynamic convex hull computation
employed as a key subroutine~\cite{BJ00,Chan99}.  
The current best time complexity (randomized expected time)
is $O( g_k(n) \alpha(g_k(n)) \log n)$ time 
given by Har-Peled~\cite{HP00}, where $\alpha()$ is the 
inverse Ackermann function. Thus, the time complexity
is very close to $O(g_k(n) \log n)$. 

However, we often need a compact ``outline'' of 
a trajectory of a parametric problem by using  
a small number of characteristic points on it.
Such an outline, generally speaking, 
will be useful as a compact data    
to control parametric problems, and possibly utilized in 
designing kinetic data structures~\cite{BGH97,AEGH98}.
Local peaks in the trajectory are considered to be 
natural characteristic points.
A key observation to investigate the $k$-level is   
that it has at most $2k-1$ local peaks 
(at most $k$ maximal peaks and at most 
$k-1$ minimal peaks)~\cite{AACS98,Eppstein95}. 
One interesting problem is to compute all the 
local peaks efficiently.
This enables us to give a decomposition of the $k$-level into 
monotone chains, and hence create an outline with a size $O(k)$ 
of the $k$-level. 

We give an algorithm to compute all the local peaks in 
the $k$-level of an arrangement of $n$ lines in 
$O( n \log n) + \tilde{O}((kn)^{2/3})$ time,
where $\tilde{O}$ is 
the big-O notation ignoring polylogarithmic factors.
The current estimate for the 
polylogarithmic factor of the second term is less 
than $\log^{5} n$; however
we do not give it explicitly in this paper, 
since it is probably loose and 
will confuse readers.
The time complexity is better than  
$O(g_k(n) \log n)$    
for some restricted range of $k$ even if 
the current lower bound of $g_k(n)$ by T\'{o}th~\cite{Toth00} 
is tight.   
If we substitute the current $O( k^{1/3} n)$ upper bound 
to $g_k(n)$, the time complexity is better than 
$O(g_k(n) \log n)$ 
if $k = O( n / \log^{c} n)$, where $c$ 
is a suitable constant. 

Another interesting question is how fast we can compute 
$\tau$ largest maximal peaks for $\tau \le k$. 
If $\tau = 1$, Roos and Widmayer~\cite{RW94} 
gave a neat method to compute the maximum 
point in the $k$-level in 
$O(n \log n + (n-k) \log^2 (n-k))$ time by using 
an efficient slope selection algorithm. 
We can compute $\tau$ largest peaks in 
$O( n \log ^2 n) + \tilde{O}((\tau n)^{2/3})$ time by combining 
Roos and Widmayer's technique and the above mentioned method
for computing all the peaks.

Finally, we investigate whether we can 
analogously treat some parametric matroids: 
Compute peaks in the trajectory of the largest 
element in the minimum weight base of a parametric matroid.
In particular,  
the graphic matroid is of wide 
interest~\cite{Eppstein95,G79,KIT95}: 
Consider a weighted undirected connected graph $G(x)$ 
with $k$ nodes and $n$ edges, such that each edge 
has a parametric weight that is linear in a parameter $x$. 
Here, $k$ and $n$ become the rank and size of the 
graphic matroid, respectively. 
Let $T(x)$ be the minimum weight spanning tree of $G(x)$ and 
consider the longest edge $e(x)$ in $T(x)$.  
Note that the minimum weight spanning tree becomes
a spanning tree that minimizes the length of the longest edge.
We call the edge $e(x)$ the {\em spanning bottleneck edge}(SBE), 
and write $SBE(x)$ and $w_{SBE}(x)$ for $e(x)$ and its weight,
respectively. 
The naming comes from the fact that  
$w_{SBE}(x)$ is the minimum value of $w$ such that 
the subgraph of $G(x)$ consisting of  
edges with weights less than or equal to 
$w$ is connected.

The following problems are important in 
sensitivity analysis:  
(1). Compute the maximum value 
and the minimum value
of $w_{SBE}(x)$ for $x \in I$, where $I$ 
is a given interval. 
(2). Compute all peaks of the trajectory
$y = w_{SBE}(x)$. 

For example, imagine a system represented by the graph 
$G$ where a link represented by 
an edge with a weight larger than a 
(controllable) threshold value 
becomes unreliable, and  
the edge weight depends on a parameter $x$ linearly within 
an interval $I$.  
For a given subinterval $J \subset I$,
we want to know the threshold value of the edge weights 
so that the graph remains connected for every  
$x \in J$.
This can be reduced to the problem (1). 
Moreover,
if we have computed all peaks in $I$ as a preprocessing
(problem (2)), we can efficiently query for
the threshold value, provided that we 
have an efficient method (shown in Section 4) 
to compute $w_{SBE}(x)$ at endpoints of $J$.
   
Both of problems (1) and (2) can be solved by computing the 
whole picture of the transitions of the minimum spanning trees, 
and the time complexity of the computation  
is $O(kn \log n)$ by using an algoirthm given by 
Fernandez-Baca et al.~\cite{FSE96}.
Note that $O(h_k(n) n^{2/3} \log^{3/4})$ 
time output-sensitive algorithm 
of Agarwal et al.\cite{AEGH98}
is better for some range of $k$, where 
$h_k(n)$ is the number of of transitions 
of the minimum spanning tree.  An 
$O(k^{1/3} n)$ upper bound~\cite{Dey97} and an 
$\Omega(n \alpha(k))$ lower bound~\cite{Eppstein95} 
are known for $h_k(n).$

Roos and Widmayer's method can be directly 
applied to the first problem.
By using dynamic maintenance algorithms~\cite{EGI97} 
of a minimum spanning tree, the time complexity 
becomes $O(\sqrt{k} n \log n)$.  
Combined with range searching techniques,
we improve the time complexity to 
$\tilde{O}( n^{8/7}k^{1/7} + n k^{1/3})$. 
We give some discussion on the second 
problem, although theoretical improvement 
on the $O(kn \log n)$ time method remains open.

\section{Preliminaries}
\subsection{Roos and Widmayer's algorithm}   

Given a set $\H$ of $n$ lines in the $x$-$y$ plane,
let ${\L}_k$ be 
the $k$-level of the arrangement of $\H$.
Let $\p$ be a point on ${\L}_k$ that has the  
maximum $y$-value $y_{max}$.  
Without loss of generality, we assume that such a point is unique. 
For any given value $\alpha$, one can decide
whether $y_{max} \ge \alpha$ or not in $O(n \log n)$ time:
We sweep on the line 
$h: y=\alpha$ from the leftmost intersection point to the right
to compute the levels of all intersection points 
on $h$ with lines in the 
arrangement. If all intersection points are above the $k$-level, 
$\alpha > y_{max}$; otherwise, $\alpha \le y_{max}$.  
By using this decision method,
a binary search algorithm works to compute $\p$, and a weakly polynomial
time algorithm with a time complexity
$O(n \log n \log \Gamma)$ can be obtained, if each coefficient
of the lines is a quotient number of integers 
with $\log \Gamma$ bits.  
Roos and  Widmayer\cite{RW94} 
applied an efficient slope selection method \cite{CSSS89} to 
transform the binary search algorithm into strongly polynomial, 
and gave an $O( n \log^2 n)$ time algorithm.
They further  
improved the time complexity to 
$O( n \log n + k \log ^2 k)$ 
for computing the minimum and 
$O( n \log n + (n-k) \log ^2 (n-k))$ for computing the maximum.

\subsection{Range query and 
Matou\v{s}ek's point set decomposition}

We use well-known (although sophisticated) 
simplex range query data structures~\cite{Aga}: 
We construct a data structure for a set $S$ of $n$ points 
in a plane such 
that given a query halfplane $H$, we can compute 
the number of points in $S$ located in $H$ efficiently. 
If we spend $O(m)$ time for constructing the data space for 
$n \log n \le m \le n^2$,  
the query time is $\tilde{O}( n/ m^{1/2})$.    
The query can be done in polylogarithmic time by using
$O( n/ m^{1/2})$ processors. 
The data structure uses $\tilde{O}(m)$ space, 
although we do not discuss space complexity in this paper.   
Moreover, we can query the number of points in 
the intersection of two (or three) halfplanes 
in the same query time if 
we ignore a polylogarithmic factor.
We can also do reporting query by spending additional 
$O(N)$ time if the region contains $N$ points.

Given a set $\H$ of $n$ lines in a plane, 
we consider the set $\D(\H)$ of 
their dual points: The dual point of a line 
$y = ax -b $ is $(a, b)$.
We construct a range searching data structure for 
$\D(\H)$. 
Given a point $p= (x_0, y_0)$, the 
set of dual points of lines 
below $p$ is the set of points in $\D(\H)$ located 
below the line $ Y = x_0 X - y_0$, where $X$ and 
$Y$ correspond to coordinates of the dual plane.   
Thus, we can compute the level of the point $p$ in the 
arrangement of $n$ lines by using half-plane range searching.
Moreover, 
we have the highest line below $p$ in the same query time. 
Also, we can 
query the number of lines which lie below 
both of a pair of query points.

A main building block for the range query is 
the point set decomposition 
structure of Matou\v{s}ek~\cite{M91}, which we also need to 
utilize directly (we only describe its two-dimensional version): 

\begin{theorem}[Matou\v{s}ek]
Given a set $S$ of $n$ points in the plane, 
for any given $r < n$, 
we can subdivide $S$ into $r$ disjoint subsets 
$S_i$ $(i=1,2,\ldots, r)$ such that 
$|S_i| \le 2n /r$ satisfying 
the following condition: Each $S_i$ is enclosed in a triangle
$\sigma_i$, and any line in the plane cuts at most 
$c r^{1/2}$ triangles among $\sigma_1,\sigma_2,\ldots,
\sigma_r$ where $c$ is a constant independent of $n$ and $r$. 
Such a decomposition can be constructed in $O( n \log n)$ time.
\end{theorem}

\section{Computing all peaks in $k$-level}

We assume $k \le n/2$ for simplicity from now on; 
if $k > n/2$, replace $k$ by $n-k$ and  
exchange maximal and minimal in the statements.
A key observation for the $k$-level is that it 
is a subset of a union of $k$ concave chains
such that all concave vertices of the 
$k$-level are vertices of these 
concave chains~\cite{AACS98}; thus, 
a $k$-level has 
at most $k$ maximal peaks and $k-1$ minimal peaks. 
We want to compute all the local peaks
in a given interval $I$ of the $x$-coordinate value.

We apply a version of 
parametric search paradigm~\cite{Me83,Salowe97}. 
However, before 
applying the parametric search, we start with a 
simpler ``$k$-branching binary search'' method.
Without loss of generality, we assume that no line in 
the arrangement is horizontal nor vertical.

We prepare two key-subroutines: {\em one-shot query} and 
{\em peak counting}: 
Let $\p(x_0)$ be the point on the $k$-level at the 
$x$-coordinate value $x_0$.
Let $\ell_k^- (x_0)$ (resp. $\ell_k^+ (x_0)$) 
be the line in the $k$-level at 
the $x$-coordinate value $x_0 - \epsilon$ (resp. $x_0+ \epsilon$)
for infinitesimally small $\epsilon > 0$.
If $x_0$ is not an $x$-coordinate value 
of a vertex on the $k$-level, 
$\ell_k^-(x_0) = \ell_k^+ (x_0)$. 
The above operation to compute the point 
(together with lines containing the point) on ${\cal L}_k$ at 
a given $x$-coordinate value is called {\em one-shot query}.
One-shot query is an analogue of 
{\em ray shooting}~\cite{Aga}, and thus the 
following lemma is basically well-known.
The complexity $q(n,m)$ given in the lemma 
is called {\em one-shot query-time} for the $k$-level:

\begin{lemma}
If we preprocess the lines in ${\cal H}$ with $O(m)$ time 
for $n \log n < m < n^2$, 
given an $x$-coordinate value $x_0$, 
we can compute $\p(x_0)$, 
$\ell_k^{-}(x_0)$ and $\ell_k^{+}(x_0)$ in 
$q(n,m) = \tilde{O}(n/m^{1/2})$ time, and also  
in polylogarithmic time by using 
$O( n/m^{1/2})$ processors. 
\end{lemma}
\label{lem:one-shot}
\begin{proof}
By using the method given in the preliminary section, 
we can compute the level of any given point 
$(x_0, y_0)$ in polylogarithmic time by 
using $O( n/m^{1/2})$ processors.  
We now apply parametric searching to have the sequential
time bound to compute the point $\p(x_0)$.
A parametric searching algorithm is 
usually stated as a sequential algorithm; however, 
it is naturally 
a parallel algorithm if 
we use a parallel decision algorithm and 
also a parallel sorting algorithm.
We remark that we can do it easier 
without using the parametric searching if we examine 
the range searching method in precise; however, 
we omit it in this paper.
\end{proof}

The peak-counting is a routine to  
compute the number of peaks 
of the $k$-level in a given  
interval $J = [x_0, x_1]$ of $x$-coordinate values efficiently. 
The following elementary lemma is essential:

\begin{lemma}
Let $f(x_0)$ and $f(x_1)$ are numbers of 
positive-slope lines below or on the $k$-level  
at $x_0$ and $x_1$, respectively.
Then, the number of maximal peaks of ${\cal L}_k$ 
in the interval $J$ is $f(x_0) - f(x_1)$. 
\end{lemma} 
\label{lem:comb}
\begin{proof}
At-most-$k$-level 
(the part of the arrangement below $k+1$-level)
is a union of 
$k$ concave chains such that all concave 
peaks in the chains appear in the $k$-level~\cite{AACS98}. 
If a concave chain among them has a peak in $J$, 
the slope of the chain must be changed from 
positive to negative. Thus,
the number of 
maximal peaks within $J$ is the difference between 
the numbers of positive slope
lines at two endpoints.
\end{proof}

We remark that $f(x_0)-f(x_1)$ equals the number of 
positive slope lines intersecting the segment 
between $\p(x_0)$ and $\p(x_1)$ if the segment has a 
nonpositive slope.

\begin{lemma}
For a given interval $J$ of $x$-coordinate value, 
the number $\kappa(J)$ of  
peaks of ${\cal L}_k$ in $J$ 
can be computed in $O(q(n,m))$ time if 
we preprocess the lines with $O(m)$ time.
Also, the number of maximal peaks can be computed in \\
$O(q(n,m))$ time.
\end{lemma}
\label{lem:count}
\begin{proof}
If we construct the dual of range search data structure 
for the set of lines with positive slopes, the number 
of positive slope lines below a given point 
$(x_0, y_0)$ can be computed 
in $O(q(n,m))$ time. Hence, $f(x_0)$ and $f(x_1)$ 
can be computed in $O(q(n,m))$ time, and the number 
of maximal peaks can be computed  
by using Lemma~\ref{lem:comb}.
The number of minimal peaks is easily computed from 
the number of maximal peaks and slopes of the 
$k$-level at endpoints.
\end{proof}

Now, we can apply a binary search paradigm to design a 
weakly-polynomial time algorithm.
First, for the input interval $I$, we compute the number of 
peaks $\kappa = \kappa(I)$ within the interval.
The time complexity for this initialization is negligible, and 
obviously $\kappa \le 2k-1$.
Next, we construct a data structure for the one-shot query 
in $O(m)$ time, where the choice of $m$ will be explained later. 
Suppose that coefficients of the equations of lines
are quotient numbers of $\log \Gamma$ bit integers.
We apply $\kappa$-branching
binary search to find all peaks; At each 
stage of the binary search, we have at most $\kappa$ 
subintervals which has at least one local peak 
of $\L_k$ (such a subinterval is called an 
{\em active interval}), 
and we recursively search 
in active subintervals.  
Thus, after examining $\kappa \log \Gamma$ candidates  
of $x$-coordinate values, we can find all of the peaks.   

\begin{proposition}
All the local peaks of ${\cal I}_k$ in the interval $I$
can be computed in $O( \kappa q(n,m) \log \Gamma)$ time. 
\end{proposition}

To make the complexity into strongly polynomial, 
we apply parametric search by using the parallel algorithm
for the one-shot query given in Lemma~\ref{lem:one-shot}
as its {\em guide algorithm}. 
We run the guide algorithm without inputting the parameter 
value (in our case, an $x$-coordinate value), and 
decide the $x$-coordinate values of the peaks
by using the sequential one-shot query 
and the counting algorithm 
of Lemma~\ref{lem:count} as {\em decision algorithms}.
The counting algorithm needs two values of 
the parameter, which corresponds to endpoints of 
each of intervals obtained by splitting active intervals
in the parametric searching process.  
Usually, parametric searching method solves 
optimization problems on monotone or convex 
functions.
Here, $k$-level is neither monotone nor convex, 
but it consists of $\kappa$ monotone fragments.
Thus, while running the guide algorithm, there are 
at most $\kappa$ different critical parameter values
to determine all
the comparisons in the current parallel step that are
necessary to proceed into the next parallel step.  
In precise, the number of different choices is the number 
of active intervals obtained breaking $I$ by the critical 
parameter values found so far in the guide algorithm.
We make a clone of the guide algorithm for each
active interval.
If the current interval is split into 
$f$ active subintervals, $f-1$ new clones are created. 
Naturally, we create at most $\kappa$ clones in our process. 
There is only one critical parameter value 
to determine the comparisons in a ``usual'' parametric search,
and such a value can be found if we run the 
decision algorithm  $O(\log N)$ times 
if the guide algorithm is a parallel algorithm on $N$ processors. 
In our case, we run the decision algorithm 
$O(\kappa \log N)$ times at each level.
Thus, we obtain the following theorem:

\begin{theorem}
All the peaks on $\L_k$ within an interval $I$ can be 
computed in 
$O( n \log n) + \tilde{O}((\kappa n)^{2/3})$ 
time if $I$ has $\kappa$ local peaks.
\end{theorem}  
\begin{proof}
The parametric searching method gives 
$\tilde{O}(\kappa q(n,m) )$ time complexity apart from the 
$O(m)$ preprocessing time. 
We balance $\kappa q(n,m) = \tilde{O}(\kappa n / m^{1/2})$ 
and $m$ to have $m = (\kappa n)^{2/3}$. 
If $(\kappa n)^{2/3} < n$, we instead use $m=n \log n$.  
This gives the time complexity.
\end{proof}

Since $\kappa \le 2k$, we have the following:

\begin{corollary}
All the peaks on $\L_k$ can be computed in 
$O( n \log n) + \tilde{O}((k n)^{2/3})$ time. 
\end{corollary}

\subsection{Computing selected peaks} 

When $\kappa$ is large, it may be too expensive to compute all 
the local peaks. 
Suppose that we want to compute $\tau$ largest 
maximal peaks in the input interval $I$ 
for $\tau \ll \kappa$ more efficiently than
computing all the peaks. 
This can be done by combining Roos and Widmayer's  algorithm 
and the algorithm given above.

We first run a binary search process 
with respect to $y$-coordinate values similarly to 
Roos and Widmayer's algorithm. 
At the intersection of the arrangement with a horizontal 
line $y = y_0$, 
we compute intervals on the line  
which are below the $k$-level 
(we call them semi-active intervals) 
in $O(n \log n)$ time. 
We next compute the sum $s(y_0)$ 
of numbers of local peaks in the semi-active intervals.
We could apply our counting method
of Lemma~\ref{lem:count}
for each semi-active intervals to compute the sum of maximal
peaks in the intervals by using $O( \tau q(n,m))$ time. 
More simply, 
we can compute it in $O( n \log n)$ time 
by counting the number of intersecting positive
slope lines with the horizontal line 
during the sweep. 
In precise, we also need information of the arrangement
at endpoints of the input interval $I$ 
if one (or both) of them 
is below $k$-level (we omit details).

If $s(y_0)$ is not between $\tau$ and $ 2\tau$, 
we continue binary search on $y_0$:  
If $s(y_0) > 2\tau$, we increase $y_0$ while 
if $s(y_0) < \tau$, we decrease $y_0$.
Thus, we can eventually find a value $y_0$ such that     
$\tau \le s(y_0) \le 2\tau$. We have spent $O( n \log ^2 n)$ time so far.
Now, we search all peaks in the union of active intervals 
by using the method given in the 
previous section. 
The following theorem is easy to see:

\begin{theorem}
We can compute $\tau$ largest maximal peaks of $\L_k$ in an interval $I$ 
in $O( n \log^2 n )+\tilde{O}((\tau n)^{2/3})$ time.  We can 
also compute $\tau$ largest local peaks
(including both maximal and minimal peaks) in the 
same time complexity.
\end{theorem}

Note that if we only use Roos and Widmayer's algorithm
in a naive fashion  
to find $\tau$ largest peaks, it would cost 
$O(\tau n \log ^2 n)$ time.   
Analogously, we can compute the 
$\tau$-smallest minimal peaks.

\begin{theorem}
We can compute $\tau$ smallest minimal peaks 
of $\L_k$ in an interval $I$
in $O(n \log^2 n) + \tilde{O}((\tau n)^{2/3})$ time.  We can 
also compute $\tau$ smallest local peaks
(including both maximal and minimal peaks) in the 
same time complexity.
\end{theorem}
\label{thm:small}

For this $\tau$ smallest peak finding 
problem, we can modify
Roos and Widmayer's method~\cite{RW94} to compute them in 
$O(n \log^2 n + k \tau \log^2 k )$ time.
Although $n + (\tau n)^{2/3} \le 2( n + k \tau) $ always holds, 
the time complexity is better than the one in 
Theorem~\ref{thm:small} by a polylogarithmic factor if   
$n^{1/2} \log^{-c} n < k < n^{1/2} \log ^{c} n$ for  
some constant $c$. Moreover, the algorithm does
not use range search data structure, and hence much 
simpler and uses smaller data space.

\begin{proposition}  
The $\tau$ smallest local peaks 
in an interval $I$ can be found in 
$O(n \log^2 n + k \tau \log^2 k )$ time.
\end{proposition}
\begin{proof}
First we search for a horizontal line $h$ such that 
it intersects $k$-level, 
and the number of peaks below it 
is not less than $\tau$.
Such a line $h$ can be found in $O(n \log ^2 n)$ time. 
Next, let $v$ and $w$ be 
the leftmost intersection and the rightmost intersection 
with the $k$-level on $h$, and let $J$ be the interval between them. 
If the $k$-level at an endpoint 
of the input interval $I$ is below $h$, 
we connect the point on $k$-level at the $x$-value of the 
endpoint to $J$ with a segment to form a chain $C$
with at most three segments. Let $\H_0$ be the set of lines in the arrangement 
intersecting with the chain $C$.
The cardinality of $\H_0$ is at most $2k$ because of Lemma~\ref{lem:comb}.
Finally, we find all peaks below $h$ by using 
the $\tau$-branching binary search.
Here, by using the windowing method of \cite{RW94},  
we should only take care of lines in $\H_0$ together with lines below endpoints of the chain $C$. 
There are at most $4k$ such lines.  
Hence, this second step can be done in $O(k \tau \log^2 k )$ time.
\end{proof}
 
\section{Bottleneck edge length in a \\ parametric  
spanning tree}

Next, we consider the parametric spanning tree problem.
Consider a connected graph 
$G = (V, E)$ with $k$ nodes and $n$ edges. 
Because of the connectivity, $k-1 \le n \le k(k+1)/2$. 
For each edge $e \in E$, we associate a weight 
function $w_e(x)$, which is linear on a parameter $x$. 
We assume that the arrangement generated by 
lines $y = w_e(x): e \in E$ is simple, i.e., no three lines
intersect at a point. We can remove this assumption by giving
a symbolic perturbation.
$G$ is denoted by $G(x)$ if it is considered as a weighted 
graph with parametric weights.
For a given value $x$, we consider the minimum spanning tree
$T(x)$ of $G(x)$.

It is known that the number of transitions of the structure of 
the minimum spanning tree  $T(x)$ is $O( k^{1/3} n)$, and
all the transitions can be computed in 
$O( kn \log n)$ time~\cite{FSE96}.   
Moreover, the average edge weight in the minimum spanning tree
is a concave function in $x$, and the value of $x$ maximizing 
the average edge weight of $T(x)$ can be computed in
$O(n \log n)$ time~\cite{FSE96}. 

As parametric matroid problems, the average edge weight 
is a counterpart of the average of $y$-values of 
$k$ lines below (or on) the $k$-level. 
A natural counterpart of the $k$-level itself in 
the minimum spanning tree is the longest
(i.e. maximum weight) edge in the minimum spanning tree.
The edge is also called the 
{\em spanning bottleneck edge} at $x$ ($SBE(x)$ in short), 
and its weight is denoted by $w_{SBE}(x)$. 
It is easily observed that $W_{SBE}(x)$ 
is the minimum value of $w$ such that  
the subgraph of $G(x)$ constructed from the  
set of edges whose weights are less than or equal to $w$ 
is connected.

It is natural and important problem 
in sensitivity analysis~\cite{G80} 
to trace the trajectory 
$y= w_{SBE}(x)$ of the weight of $SBE(x)$. 
Analogously to the $k$-level, there are at most $k$ 
maximal peaks in the trajectory $y = w_{SBE}(x)$.
We want to compute peaks in the trajectory.

\subsection{One-shot query for the longest edge in MST}

We first consider efficient query for $SBE(x_0)$
at any given value of $x_0$ of the parameter. 
This query is called {\em one-shot query for the $SBE$}.
A naive method is the following: First construct $T(x_0)$ in 
$O(n)$ time, and select its longest edge.   
Instead, we use the  Matou\v{s}ek's set partition.
In the dual space, the dual points of $n$ weight functions
of the edges are partitioned into $r$ subsets of size 
$O(n/r)$. Each subset is contained in a triangle, and 
$O(r^{1/2})$ triangles are cut by any query line.

Accordingly, we partition the set of $n$ edges
into $r$ subsets each has $O(n/r)$ edges. 
For each subset, we compute a spanning forest 
(irrelevant to edge weights) and store the connected components
except singletons. Thus, each component has a forest with 
$O(\min\{ k, n/r\})$ edges.
This computation can be done in $O(n)$ 
additional time.

If we are given a parameter value $x_0$,  
we sort $O(r)$ vertices of the 
triangles with respect to the inner product 
of them with the vector $(x_0, 1)$.  
We do binary search on this sorting list.
We guess a vertex $v$, and consider a line 
$\ell:  Y = x_0 X + c$ which goes through $v$. 
We recognize the triangles which are below $\ell$; 
thus, the edges in the subsets associated with 
the triangles has weights less than $c$.   
We construct a spanning forest $F$ of the union 
of forests in these subsets: 
since they have $O(rk)$ edges, this can be done 
in $O( rk )$ time. 
If the forest $F$ is a spanning tree, we decide that 
$v$ is too large in the sorting list, and 
continue the binary search.

Otherwise, we consider the subsets associated with 
the triangles cut by $\ell$. 
They contain $O( n/ r^{1/2})$ points in total. 
We sort them with respect to the weights, and 
greedily insert them into $F$ until we have a spanning tree.
If we do not have a spanning tree, we decide $v$ is 
too small, and continue the binary search. 
If we have a spanning tree, 
we decide $v$ is a candidate, but it may be too large,
and continue to search for  
the lowest vertex $v$ satisfying the above condition, and 
return the longest edge in the tree for that $v$. 
Note that the spanning tree is not  
a minimum spanning tree in general; however, we correctly
recognize the longest edge in a minimum spanning tree.

This process needs $O( rk + n/r^{1/2})$ time, and 
we do this process $O(\log r)$ times during the binary search.
Thus, if we set $r = (n/k)^{2/3}$,  
the time complexity is $O(n^{2/3}k^{1/3} \log (n/k))$, 
which is slightly better than $O(n)$ if $k = o(n)$. 
By applying a hierarchical subdivision, 
we can further improve it: 
We fist start $r = r_1$, and 
decompose the subset of size $O(r_1n)$ into $r_2$ smaller 
subsets, where $r_2 = r_1^{1/2}$, and we further continue 
the refinement for $r_i = r_{i-1}^{1/2}$ until 
$r_i$ becomes below a constant. 
The query time becomes \\ 
$k (r_1 + r_1^{1/2} r_2 + \ldots + 
(r_1r_2 \ldots r_{i-1})^{1/2} r_{i}) + 
n/(r_1r_2 \ldots r_i)^{1/2}$. 
Setting $r_1 = (n/k)^{1/2}$, 
this enables $\tilde{O}((nk)^{1/2})$ 
time computation for $SBE(x_0)$. 
Similarly to the case of halfspace range searching,
we can combine hierarchical cutting~\cite{Chazelle93} 
of the arrangement to have a 
preprocess-query trade-off (we omit details in this version).
Indeed, we have the following proposition:
  
\begin{proposition}
If we spend $\tilde{O}(m)$ preprocessing time for 
$n \le m $, we can do the one-shot query for 
$SBE$ in \\
$\tilde{O}( n/ (m/k)^{1/2} + k)$ time.  
\end{proposition}

Moreover, we will later use the following 
{\em two-shot reporting query} for a spanning forest, 
which reports 
a spanning forest consisting of edges whose weight 
functions are below both of given two query points 
$(x_0, y_0)$ and $(x_1, y_1)$. 
This can be done similarly to one-shot query
(this is a counterpart of the simplex range searching
if the one-shot query is a counterpart of the halfplane 
range searching).
  
\begin{proposition}
If we spend $\tilde{O}(m)$ preprocessing time for 
$n \le m $, we can do the two-shot 
reporting query in \\
$\tilde{O}(n/ (m/k)^{1/2} + k )$ time.  
\end{proposition}

\subsection{Computing the maximum peak}

Let us consider the problem of computing the 
maximum peak in $I$. 
First, we straightforwardly apply Roos-Widmayer's algorithm.
For a given $y$-value $y_0$, we want to decide whether 
$Max_{x \in I} w_{SBE}(x) \le y_0$ or not.
We dynamically update
the spanning forest associated with edges with weight 
below $y_0$ from $x= x_0$ to $x= x_1$ if $I = [x_0, x_1]$. 
If we find a value $x \in I$ such that 
the spanning forest becomes a spanning tree, we know  
$Max_{x \in I} w_{SBE}(x) \le y_0$.
It costs $O(k^{1/2})$ time to update a minimum spanning forest
for insertion and deletion of edges.
Suppose that we sweep on the line $y= y_0$ 
updating the minimum spanning forest.
The line $y=y_0$ has at most $n$ intersections with 
lines associated with weight functions, and hence 
the method needs $O(n k^{1/2})$ time for the decision.
Thus, the maximum peak can be found in 
$O(n k^{1/2} \log n)$ time.

We try to improve the above time complexity. 
We subdivide the line $y=y_0$ into $\lceil n/s \rceil$ intervals 
such that each interval contains at most $s$ intersection
points. 
For each interval $I_i = [x_i, x_{i+1}]$, 
we perform the two-shot reporting query 
at $(x_i, y_0)$ and $(x_{i+1}, y_0)$. 
The reported forest $F$ is constructed from 
edges whose weight is less than $y_0$ both at $x=x_i$ and
$x= x_{i+1}$. If the forest has more than $s+1$ connected 
components, it is impossible that $w_{SBE}(t) \le y_0$ 
for a $t \in I_i$. 
Otherwise, we dynamically maintain the spanning tree, 
where we contract nodes into at most $s+1$ super nodes 
each of which associate with a connected component of 
the forest $F$.
Our graph has only $s$ edges, and hence the update 
can be done $O(\sqrt{s})$ time per intersection.

Hence, total time complexity becomes \\
$\tilde{O}( n \sqrt{s} + (n/s)[n/(m/k)^{1/2} + k] + m)$. \\
If we optimize this, we have 
$\tilde{O}( n^{8/7} k^{1/7} + n k^{1/3})$. 
This is an improvement over $O(n k^{1/2})$, 
since $n \le k(k+1)/2$.   

The minimum of $w_{SBE}(x)$ can be analogously computed.
Hence, we have the following theorem:

\begin{theorem}
For a given interval $I$ of the parameter value $x$, 
we can compute both the maximum and the minimum 
of $w_{SBE}(x)$ for  $x \in I$  in 
$\tilde{O}(n^{8/7} k^{1/7} + n k^{1/3})$ time.
\end{theorem}

We can generalize the above theorem for the {\em truncated 
matroid} of the graphic matroid to obtain the 
following proposition (we omit the proof):

\begin{proposition}
For a constant $c$, let $w_{SBE-c}(x)$ be the minimum value
of $w$ such that the set of edges in $G(x)$ with weights  
less than or equal to $w$ has at most $c$ connected
components.    
We can compute both 
the maximum and the minimum of $w_{SBE-c}(x)$ in 
$\tilde{O}(n^{8/7} k^{1/7} + n k^{1/3})$ time.
\end{proposition}


\subsection{Computing all local peaks for $SBE$ }

It is desired to apply the method of computing all peaks
in the $k$-level to $SBE$ in a parametric graph.
It is known that there are at most $k$ maximal peaks 
in $y = w_{SBE}(x)$. 
Unfortunately, we have no good method to know the 
number of peaks within an interval $I = [x_0, x_1]$ exactly,
since we do not have a property that is a counterpart of 
Lemma~\ref{lem:comb}. 
The only known method for the authors 
is to compute the minimum spanning 
trees $T(x_0)$ and $T(x_1)$ explicitly, and 
compute the difference $d(I)$ between   
the number of edges whose weight functions are 
positive slopes.
For any disjoint set of intervals, the sum of $d(I)$ 
over the intervals is at most $k$, and 
$d(I)$ gives an upper bound of maximal peaks 
of $y= w_{SBE}(x)$ within $I$.  However, it is often an 
overestimate, since $d(I)$ gives the number of maximal peaks 
of $k$ trajectories of weights of 
all edges (not only maximum one) in the parametric 
minimum spanning tree, where we only include peaks 
where an edge in the spanning tree is replaced by 
an edge outside the spanning tree. 

By using the above observation, 
we have an $\tilde{O}( k f(n,k))$ time algorithm
if we have an algorithm to compute $T(x_0)$ for 
a given $x_0$ in $O(f(n,k))$ time.
Unfortunately, we only have
$f(n,k)= O(n)$, which leads to an $\tilde{O} ( kn)$ time 
algorithm, which is inferior to a known algorithm to compute 
all transitions of the parametric minimum spanning tree.  
We can compute $\tau$ largest peaks in transitions of 
edge weights in MST in $\tilde{O}(\tau n)$ time,
if we include all peaks of all edges in MST; however, 
the number of peaks appeared at the transitions 
of the longest edge among them may be much smaller than 
$\tau$.  

Although the above method is not attractive for the minimum
spanning tree, the method is applicable 
to any parametric matroid, 
and hence it is useful if we 
do not have a dynamic algorithm 
to maintain a minimum weight base.
The current 
$O(kn \log n)$ time algorithm to compute the transitions 
of parametric minimum spanning tree needs $O(k^{2/3})$ time 
method (indeed, it can be done in $O(k^{1/2})$ time) 
to update a minimum spanning tree. Thus,
the analogue of the algorithm 
needs $\tilde{O}( k^{1/3} n q  + kn)$ time to compute 
all the transitions of a parametric matroid of rank $k$, where
$q$ is the time complexity to update its minimum weight base. 

\section{Concluding remarks}
The number of maximal peaks in a $k$-level is 
known to be at most
 ${}_{k} C _{d}$ (number of combinations choosing 
$d$ elements from $k$ elements) if we have $d$ dimensions
\cite{Clarkson}.
Hence, this is much smaller than the complexity of 
whole arrangement, especially if $k$ is much smaller than $n$.  
However, to the author's knowledge, 
the problem of computing peaks in the $k$-level 
for a higher dimensional arrangement is open. 
Although the algorithm of Roos and Widmayer can be applied 
to $3$-dimensional case, it needs $\tilde{O}(n^{2})$ time
to compute the largest peak (i.e. global maximum) if 
we naively implement it.
One necessary constituent is to develop a counterpart of 
Lemma~\ref{lem:count}: Given 
an arrangement of $n$ hyperplanes in the 
three-dimensional space, preprocess it, and for any given  
three points $A$, $B$, and $C$ in 
the plane $z=0$, decide whether the triangle  $ABC$ contains 
(a projection) of a peak in the $k$-level or not efficiently.
For the purpose, we probably need 
a counterpart of Lemma~\ref{lem:comb}: 
Give a criterion of the existence of a peak from the 
information of the set of hyperplanes below $k$-level at each 
of $A$, $B$, and $C$.  
In two-dimensional space, the lines are classified into 
positive slope lines and negative slope lines, while 
this kind of natural discrete classification of planes
in the space does not exist.
Moreover, it is difficult to find a counterpart of 
concave chain decomposition for three dimensional 
$k$-level~\cite{KT99}.
These lacks make the problem difficult,
although the authors think it is quite attractive.

Another interesting problem is an an extension of  
parametric $SBE$ problem to the 
$c$-edge-connectedness for $c \ge 2$. 
Here, we hope we can develop 
efficient solutions by combining geomertic methods and 
graph theoretical methods\cite{EGI97,Fred91}.  

\subsection*{Acknowledgement}
The authors gratefully acknowledge referees of the SOCG 
conference for informing several references on 
recent improved algorithms for computing $k$-level and 
parametric minimum spanning trees.



\begin{thebibliography}{99}

\bibitem{Aga}
P. Agarwal, ``Range Searching,'' 
Section 31 of {\em Handbook of Discrete and 
Computational Geometry,} (1997), 575--598, CRC Press.

\bibitem{AACS98}
P. Agarwal, B. Aronov, T. Chan, and  M. Sharir,
``On Levels in Arrangement of Lines, Segments, Planes, and 
Triangles,'' 
{\em Discrete \& Comput. Geom.,} {\bf 19} 
(1998), pp. 315-331. 

\bibitem{AEGH98}
P. Agarwal, D. Eppstein, L. Guibas, and M. Henzinger,
``Parametric and Kinetic Minimum Spanning Trees.'' 
{\em Proc. 39th IEEE FOCS} (1998) pp.596-605.
\bibitem{BGH97}
J. Bash, L. Guibas, and H. Hershberger,
``Data Structures for Mobile Data,"
{\em Proc. 8th ACM-SIAM Symp. on Disc. Alg.,} (1997), pp. 747-756. 
\bibitem{BJ00}
G. Brodal and R. Jacob,
``Dynamic Planar Convex Hull with Optimal 
Query Time and $O(\log n \cdot \log\log n)$
Update Time''
{\em Proc. 7th SWAT, LNCS 1851} (2000),
pp. 57-70. 

\bibitem{Chan99}
T. Chan, 
``Dynamic Planar Convex Hull Operations
in Near-Logarithmic Amortized Time,''
{\em Proc. 40th IEEE FOCS} (1999) pp.92-99.

\bibitem{CSSS89}
R. Cole, J. Salowe, W. Steiger, and E. Szemer\'{e}di,
``An Optimal Algorithm for Slope Selection,''
{\em SIAM J. Comput.,} {\bf 18-4} (1989), pp. 792-810.

\bibitem{CSY87}
R. Cole, M. Sharir, and C. Yap,
``On $K$-hulls and Related Problems,'' 
{\em SIAM J. Comput.,} {\bf 16} (1987), pp. 61-77.

\bibitem{Chazelle93}
B. Chazelle,
``Cutting Hyperplanes for Divide-And-Conquer'',
{\em Discrete \& Comput. Geom.,} {\bf 9} (1993) pp. 145--158. 

\bibitem{Clarkson}
K. Clarkson, 
``A Bound on Local Minima of Arrangement
That Implies the Upper Bound Theorem,''
{\em Discrete \& Comput. Geom.,} 10 (1993), pp. 427-433.

\bibitem{Dey97}
T. Dey,
``Improved Bound on Planar K-Sets and Related Problems,'' 
{\em Discrete \& Comput. Geom.,} 19 
(1998), pp. 373-383. 

\bibitem{Edels}
H. Edelsbrunner,
{\em Algorithms in Combinatorial Geometry},
ETACS Monographs on TCS 10, Springer-Verlag, 1987.

\bibitem{Eppstein95}
D. Eppstein,
``Geometric Lower Bounds for Parametric Matroid Optimization,''
{\em Discrete \& Comput. Geom.,} (1998), pp. 463-476.

\bibitem{EGI97}
D. Eppstein, Z. Galil, G.F.Italiano, and A. Nissenzweig,
``Sparcification--A Technique for Speeding Up Dynamic
Graph Algorithms,''
{\em J. ACM,} {\bf 44-5} (1997), pp. 669-696. 


\bibitem{FSE96}
D. Fernandez-Baca, G. Slutki, and D. Eppstein,
`` Using Sparcification for Parametric Minimum Spanning Tree
Problems,'' {\em Nordic J. Computing,} {\bf 3-4} (1996), 
pp. 352-366.

\bibitem{Fred91}
G. Frederickson, 
``Ambivalent data structures for dynamic 2-edge-connectivity 
and $k$-smallest spanning trees'' 
{\em Proc. 32nd FOCS} (1991) 632--641.


\bibitem{G79}
D. Gusfield, 
``Bound for the Parametric Spanning Tree Problem,''
{\em Proc. Humbolt Conf. Graph Theory, Combinatorics, 
and Computing,}  (1979), pp. 173-183.

\bibitem{G80}
D. Gusfield,
{\em Sensitivity Analysis for Combinatorial Optimization,}
Ph. D. Thesis, Memorandum No. UCB/ERL M80/22, U.C. Berkeley, 1980.

\bibitem{KIT95}
N. Katoh, K. Iwano, and T. Tokuyama,
``On Minimum and Maximum Spanning 
Trees of Linearly Moving Points,''
{\em Discrete \& Comput. Geom.,} 13 (1995) pp. 161-176.

\bibitem{HP00}
S. Har-Peled, 
``Taking a Walk in a Planar Arrangement''
{\em SIAM J. Comput.} {\bf 30}
(2000) pp. 1341-1367.


\bibitem{KTT99}
N. Katoh, H. Tamaki and T. Tokuyama,
``Parametric Polymatroid Optimization and Its Geometric Applications,''
{\em Proc. 10th ACM-SIAM SODA,} (1999), pp. 517-526. 

\bibitem{KT99}
N. Katoh and T. Tokuyama, 
``Lov\'{a}sz's Theorem for the $K$-level of an Arrangement of Concave 
Surfaces and Its Applications," {\em Proceedings 
of 40th IEEE FOCS,} (1999), pp. 389-398.

\bibitem{L71} 
L. Lov\'{a}sz, 
``On the Number of Halving Lines,'' 
{\em Ann. Univ. Sci. Budapest, Et\"{o}v\"{o}s, Sect. Math., 14}, 
(1971) pp. 107-108.   

\bibitem{M91}
J. Matou\v{s}ek, 
``Efficient Partition Trees,'' 
{\em Discrete \& Comput.
Geom.,} {\bf 10 }(1992), pp. 315--334.  

\bibitem{Me83}
N. Megiddo, ``Applying Parallel Computation Algorithms
in the Design of Serial Algorithms,''
{\em J. ACM,} {\bf 30} (1983), pp.852--865.

\bibitem{RW94}
T. Roos and P. Widmayer, 
``K-Violation Linear Programming,''
{\em Information Processing Letters,} {\bf 52}
(1994), pp. 109-114. 

\bibitem{Salowe97}
J. Salowe, Parametric Search,
Section 37 of {\em Handbook of Discrete and 
Computational Geometry,} (1997), 683--695, CRC Press.

\bibitem{Toth00}
G. T\'{o}th,
``Point Sets with Many $k$-sets,''
{\em Proc. 16th SOCG,} (2000), pp. 37-42.

\bibitem{WEB}
Web page on k-set, dissecting lines, and parametric optimization:  
http://liinwww.ira.uka.de/searchbib/Theory/kset.

\end{thebibliography}
\end{document}